# Why Aren't Protein Residue Networks Smaller Worlds

Susan Khor[1]

[1]Montreal, Canada
slc.khor@gmail.com

**Abstract**

In this paper, in silico experiments are performed to investigate why protein residue networks (PRNs), i.e. networks induced by spatial contacts between amino acid residues of a protein, do not have shorter average path lengths (APLs) in spite of their importance to protein folding. We find that shorter average inter-nodal distances does not necessarily imply better search performance, i.e. more successful protein folding. Search performance of a zero-temperature Metropolis style hill-climber was not significantly improved by randomizing only the long-range links of PRNs even though such randomization significantly reduces APLs of PRNs while retaining much of the clustering and positive degree-degree correlation inherent in PRNs. However, this result is contingent upon the optimization function. We found that the optimization function which places PRNs in a favorable spot in the space of possible network configurations considered in this paper parallels an existing view in protein folding theory that neither short-range nor long-range interactions dominate the protein folding process. These findings suggest the existence of explanations, other than the excluded volume argument, beneath the topological limits of PRNs.[1]

## Introduction

Breaking the code underlying protein folding has remained an intellectually tantalizing puzzle as well as a problem of great practical significance. Everything a protein requires for correct folding under normal circumstances appears to be embedded in its amino-acid sequence (Afinsen 1973), although a minority rely on the aid of water and chaperone molecules. Due to the large sizes that amino-acid sequences can take, a random search approach to protein folding is deemed infeasible for practical biological purposes (Levinthal 1969). However, an argument based on separability of the protein folding problem, i.e. that the problem can be separated into parts which can then be solved independently and assembled into an optimal solution[2], has been conceived as a way out of Levinthal's paradox (Zwanzig et al 1992; Karplus 1997). This argument is supported by some sections of protein sequences having a propensity to fold to their native secondary structures.

In general, protein folding is a process that occurs in stages. What essentially begins as a linear hetero-polymer (organized as a backbone with protruding side chain groups) obtains local structure in the form of secondary alpha helices and beta sheets and finally global structure as the secondary structures arrange themselves compactly in three dimensions. For a long time, this spontaneous biological self-organization has been attributed to various inter-atomic physical forces and chemical constraints impacting a protein molecule. However, in the last decade or so, another theory based on the network topology of a protein's native state has blossomed. In this other theory, a network view of protein molecules (mostly in their native states) is adopted.

The general recipe to transform a protein molecule into a network is to represent amino acid residues ($C_\alpha$ or $C_\beta$) as nodes, and contact (spatial, non-covalent) distances between pairs of amino acid residues below a certain threshold as links. Such *protein residue networks* (*PRN*) are constructed from the Cartesian coordinates of amino acid residues of protein molecules stored in the Protein Data Bank (*PDB*) (Berman et al 2000). There are variations to the general recipe however. For instance, a PRN may represent several non-homologous proteins rather than a single protein, e.g. the protein contact map in (Vendruscolo et al 2002). PRNs may also represent different aspects (e.g. surface or core), states (e.g. native or transitional), structural classes (e.g. *α, β, α+β* or *α/β*), or types (e.g. globular or fibrous) of proteins (e.g. Atilgan et al 2004). Further, the nodes and links of PRNs may carry different meanings, e.g. the atoms of the side chain group of an amino acid may be included so that a node may represent more than one atom and multiple links between nodes or weighted links are allowed (Green and Higman 2003).

By examining PRNs, researchers have compiled a list of topological characteristics shared by a diverse (in terms of structural class, homology and taxon) set of proteins and speculated on the reasons for the observed topological characteristics in relation to protein folding. A common feature of protein residue networks is their *small-world* nature, i.e. they have lattice like clustering coefficients but random graph like diameters and average path lengths (Watts and Strogatz 1998). The need for rapid communication between amino acid residues to facilitate *interaction cooperativity* crucial for protein folding is frequently cited as the reason for the small-world feature of PRNs (Vendruscolo et al 2002; Dokholyan et al 2002; Atilgan et al 2004; Del Sol et al 2006). PRNs are also reported to exhibit high assortativity values which can be related to protein folding speeds (Bagler and Sinha 2007). We discuss network characteristics of PRNs further on.

---

[1] This is an independent research paper, part of which was completed during the author's stay at Collegium Budapest, Hungary who generously provided the computer resources for most of the experiments. The author is currently a post-doctoral researcher in Montreal, Canada.

[2] For a more colourful description, see Herbert Simon's parable of the two watchmakers in *The Sciences of the Artificial*, 1969 MIT Press.

In this paper, we set out to understand why, given the assumed importance of rapid inter-residue communication to protein folding, PRNs are not smaller-worlds or equivalently do not have shorter average shortest path lengths. We address this question from a search perspective, which is not unusual given the common formulation of protein folding as a search problem. We define a spin-glass like problem on a PRN and use the performance of a local search (a hill-climber in the fashion of the Metropolis algorithm with zero probability of assuming a higher temperature configuration) to assess the effect of changes in network topology, specifically average path length (APL), on search performance. The experiments are conducted under several conditions, motivated by existing literature on protein folding theory.

## Method

### Protein residue network construction

Our *PRN* has N nodes (one for each amino acid of a protein) and M links. An undirected link is placed between a pair of nodes representing the Cα atom of amino acids when the node pair is situated less than 7Å apart from each other. The small-world property of PRNs is not overly sensitive to the choice of this threshold value (Bartoli *et al* 2007). Distance between node pairs is the Euclidean distance between their 3D Cartesian coordinates obtained from the Protein Data Bank or *PDB* (Berman *et al* 2000). The M links are partitioned into two sets: long-range links (*LE*) and short-range links (*SE*). A link between nodes *x* and *y* is classified as *long-range* if their absolute distance on the amino acid sequence chain is more than 9 (Green and Higman 2003). Long-range links connect amino acids which are distant in the primary structure but are in close spatial proximity in the tertiary structure.

### Test data set

A PRN is built for each protein in the GH64 dataset (Figures 1&2) which was selected from literature surveyed, specifically (Green and Higman 2003). The dataset encompass proteins from different protein classes, fold types and branches of life. Proteins which did not form a single connected component (i.e. 1cuk and 1ho4), or had unusually high link density (i.e. 1feo) in its PRN were excluded from the dataset. So too were proteins with more nodes in their PRN than their DSSP output (Kabsch and Sander 1983) (i.e. 2hmz and 1epf). We use the output from DSSP (Dictionary of Protein Secondary Structure) as globally optimal strings in our search problem. If the reverse situation occurs, the DSSP output is truncated. A second dataset, EVA132, is used to increase confidence of key results in this paper. The EVA132 protein dataset was extracted from the list of 3477 unique chains archived by EVA (Rost, 1999). 200 proteins were selected at random from this list, with no overlap with GH64. PRNs for these 200 proteins were constructed and selected in the same manner as GH64, yielding 132 well-formed PRNs. EVA132 PRNs possess similar network characteristics as GH64 PRNs. Detailed information on both sets can be found in http://arxiv.org/abs/1011.2222.

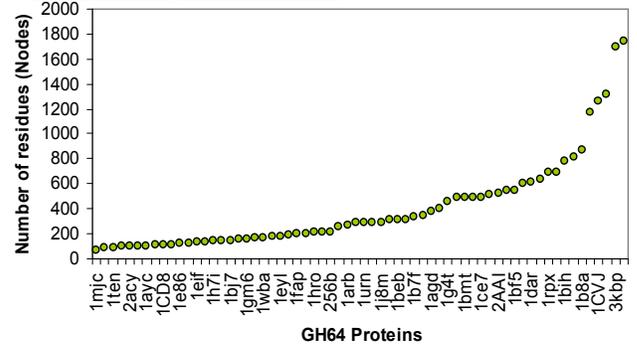

**Figure 1** Size of GH64 proteins in terms of the number of Cα atoms. PIDs are:1mjc, 1gvp, 1ten, 1ris, 2acy, 1tlk, 1ayc, 1sha, 1CD8, 1d4t, 1e86, 2fgf, 1eif, 1pdo, 1h7i, 1amx, 1bj7, 1aep, 1gm6, 3rab, 1wba, 1rbp, 1eyl, 153L, 1fap, 1nsj, 1hro, 1jr8, 256b, 1ICE, 1arb, 1vlt, 1urn, 1amp, 1j8m, 1cjl, 1beb, 1OBP, 1b7f, 1hng, 1agd, 1aye, 1g4t, 1eov, 1bmt 7tim 1ce7, 1hwn, 2AAI, 1fbv, 1bf5, 1jly, 1dar, 1eun, 1rpx, 1bbp, 1bih, 1psd, 1b8a, 1ava, 1CVJ, 3eca, 3kbp, 1dio.

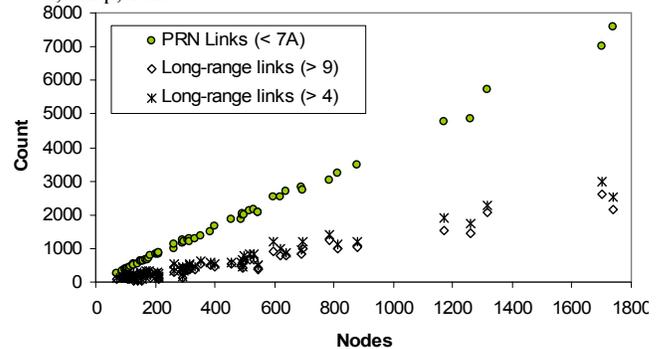

**Figure 2** Link count M, by protein size for GH64 PRNs.

### Search problem and search algorithm

We define a spin-glass like problem on a PRN and use the performance of a local search (a hill-climber in the fashion of the Metropolis algorithm with zero probability of assuming a higher temperature configuration) to assess the effect of changes in network topology, specifically average path length, on search performance. Starting at random points in a search space comprising $\{0, 1, 2\}^N$ strings[3], the problem is to find **s**, the unique globally optimal string defined by the DSSP output (Kabsch and Sander 1983) for a PRN reduced with the following rules: 0 represents H, I, and G, 1 represents E and B, and 2 represents others. The unique global optimum is reachable by maximizing the following fitness function which is derived from (Bryngelson and Wolynes, 1987):

$$\sum_{i=0}^{N} g(s_i, \mathbf{s_i}) + \sum_{i=0}^{M} f(e_i, s, \mathbf{s})$$ . Define $s_i$ as the current value of

the $i^{th}$ element in string *s*. $g(s_i, s_i) = 1$ if $s_i = \mathbf{s_i}$ and 0 otherwise.

---
[3] Incidentally, $3^N$ search spaces are common in discrete models of protein folding, e.g. 3 possible peptide bond torsion angles, and 3 possible bonds between hydrophobic (H) and polar (P) residues.

Define $e_i$ as the $i^{th}$ link in a PRN and $e_i$ connects nodes $j$ and $k$. $f(e_i, s, \mathbf{s}) = 1$ if $|s_j - s_k| = |\mathbf{s}_j - \mathbf{s}_k|$ and 0 otherwise. The $g$ term ensures a unique global optimum[4] while the $f$ term introduces frustration, i.e. the required ruggedness feature into the fitness landscape (Dill et al 1995, p.585).

The local search algorithm is a hill climber which at each time step, the value of a single randomly chosen element assumes a different value chosen randomly from {0, 1, 2}, and never moves down hill to less fit points. For each run, the hill climbing algorithm is iterated until **s** is found, or the fitness function has been evaluated 1 million times. 20 independent runs are made per PRN. A total of 1280 (64 x 20) and 2640 (132 x 20) runs are made for GH64 and EVA132 respectively.

## Network Characteristics of PRNs

**Node degree** measures the number of contacts or direct neighbors a node has in a PRN. Gaci and Balev (2009) remarked on the homogeneity of node degree in their PRN called SSE-IN which only considers secondary structure elements. The mean node degree of their SSE-INs increased very slightly with protein size and fell within the range of 5 and 8. The absence of nodes with much higher degrees is attributed to the excluded volume effect which imposes a physical limit on the number of residues that can reside within a given radius around another amino acid. The mean node degree (K) of the GH64 PRNs averages at 7.9696 with a standard deviation of 0.3126, and is independent of protein size (Figure 3).

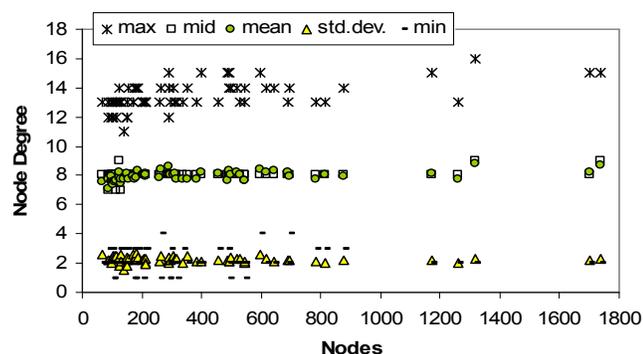

**Figure 3** Node degree summary statistics for GH64 PRNs.

**Clustering** or transitivity reflects the cliquishness of nodes in a network: if node X connects to node Y and to node Z, how likely is it that nodes Y and Z are connected to each other? A convenient way to measure network clustering is by taking the average clustering of all nodes in a network to yield the clustering coefficient as follows: $C = \frac{1}{N}\sum_{i=1}^{N}\frac{2e_i}{k_i(k_i-1)}$

where $k_i$ is the degree of node $i$, and $e_i$ is the number of links that exist amongst the $k_i$ nodes (Watts and Strogatz 1998). Independent of protein size, the $C$ values for GH64 PRNs ($C_{GH64}$) are significantly higher than $C_{RANDOM}$, and closest to

---

[4] It also has a separable or a smoothing effect; without the $f$ term, there is no interaction between variables.

---

$C_{LATTICE4}$ (Figure 4). Lattice$V$ is a linear lattice with $V/2$ nearest neighbours to the left and to the right where possible.

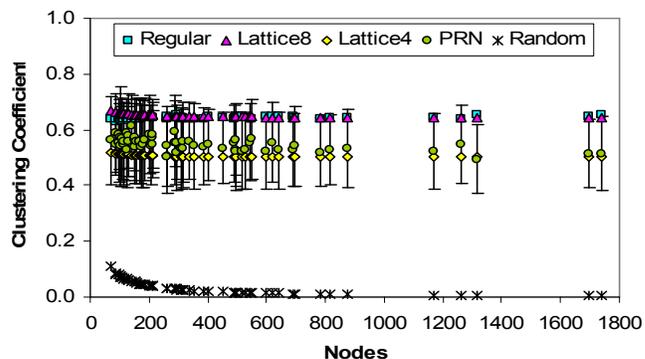

**Figure 4** $C_{GH64}$ values (± one standard deviation) compared with $C_{Lattice8}$, $C_{Lattice4}$ and the theoretical $C$ values for regular ($C_{REGULAR}$) and random ($C_{RANDOM}$) networks of the same size (same number of nodes). $C_{RANDOM} \sim K/N$, and $C_{REGULAR} = 3(K-2)/[4(K-1)]$, where K is average degree and N is number of nodes (Watts 1999). We use K=8 (see Figure 3).

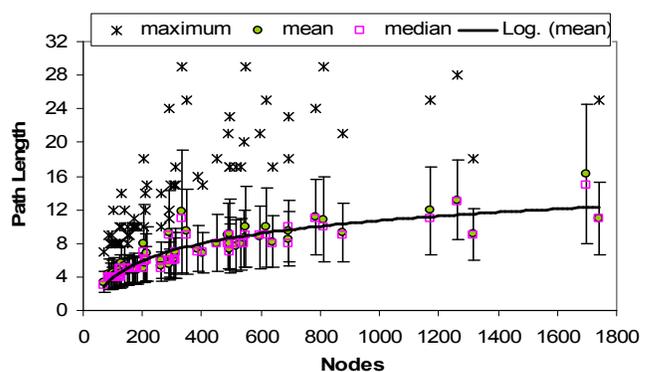

**Figure 5** Diameter, average (± one standard deviation) and median path length for GH64 PRNs.

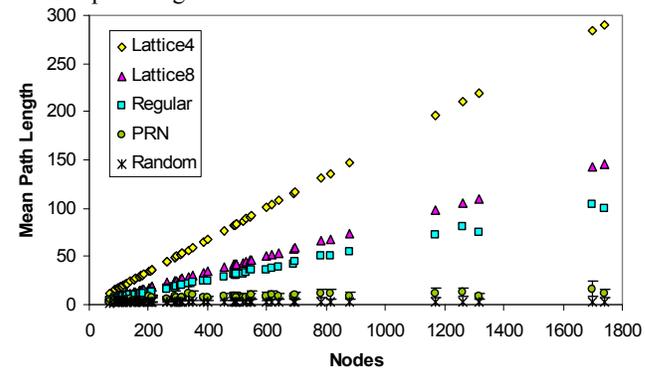

**Figure 6** APLs of PRNs are much closer to APLs of random networks ($APL_{RANDOM}$) than to APLs of regular networks ($APL_{REGULAR}$). $APL_{RANDOM} \sim \ln N / \ln K$, and $APL_{REGULAR} = N(N + K-2)/[2K(N-1)]$, where K is average degree and N is number of nodes (Watts 1999). We use K=8 (see Figure3).

The **average path length** (APL) of a network is the average length of a set of shortest paths between all node-pairs. The average path length for GH64 PRNs ($APL_{GH64}$) increases logarithmically with increases in protein size (nodes) (Figure 5). When compared with average path lengths of other canonical networks, $APL_{GH64}$ is much shorter than the average path lengths of regular graphs ($APL_{REGULAR}$) and approximate the average path lengths expected of random graphs ($APL_{RANDOM}$) of the same size (Figure 6). $APL_{GH64}$ is also much shorter than both $APL_{LATTICE8}$ and $APL_{LATTICE4}$ (Figure 6).

The **small-world property** is a combination of high clustering and short inter-nodal distances (average path length increases logarithmically with network size), two conditions that from the above exposition, GH64 PRNs satisfy.

**Assortativity** refers to the extent that nodes associate or connect with their own kind. A common form of assortativity measured for PRNs is node degree. Positive degree-degree assortativity refers to the proclivity of nodes with small (large) degree to link with other nodes of small (large) degree. Using the method in (Newman 2002), Bagler and Sinha (2007) report degree-degree correlation coefficients up to 0.58, which is considered unusual for networks with biological origins. Nonetheless, the positive assortativity values could be correlated in a positive manner to protein folding speeds (Bagler and Sinha 2007). Similarly, we find positive degree-degree correlations in the GH64 PRNs independent of protein size. The assortativity values average at 0.3387 with a standard deviation of 0.0536, which is much higher than observed for randomized PRNs (randAll) (Figure 7). For randAll networks, PRNs are randomized in the usual manner by rewiring nodes while preserving node degrees and without introducing multiple links between nodes (Maslov and Sneppen 2002).

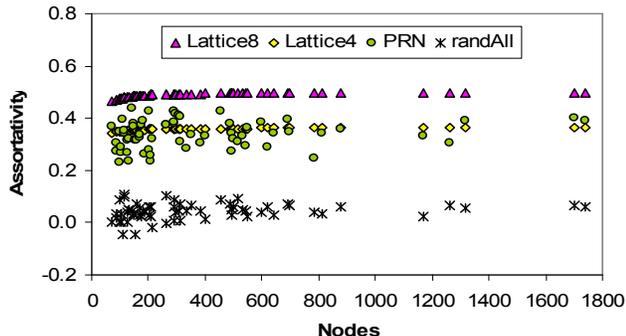

**Figure 7** The GH64 PRNs have positive degree assortativity, with values closest to those for Lattice4.

As with clustering (Figure 4), the assortativity values for GH64 PRNs are closest to Lattice4 (Figure 7). Bartoli *et al* (2007) commented that links encompassing a protein's backbone (which are short-ranged) is the main source of the relatively high levels of clustering in PRNs. We observed that short-range links (SE) are also responsible for much of the positive assortativity in PRNs. Figure 8 shows the effect of randomizing different sets of links in GH64 PRNs. Both clustering and assortativity levels show larger decreases when only the short-range links are randomized (randSE), compared with when only the long-range links are randomized (randLE). The APLs of PRNs are significantly reduced in both randSE and randLE networks. Hence it is possible, by randomizing only the long-range links, to rearrange the links of a PRN such that the APL is significantly reduced while preserving clustering and positive degree-degree assortativity coefficients at levels higher than would be in random graphs. Our question is thus: if short inter-nodal distances are important for protein folding, *why aren't the average path lengths of PRNs shorter?*

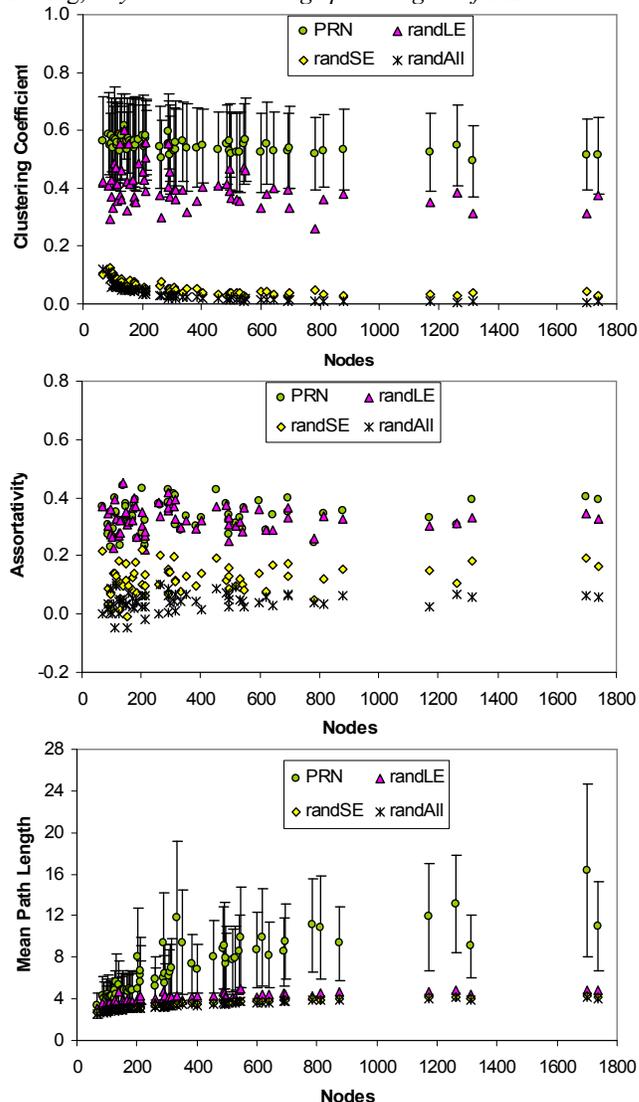

**Figure 8** Effect of randomizing different sets of PRN links on Clustering Coefficient (top), Assortativity (middle), and Average Path Length (bottom). Error bars denote one standard deviation from the mean.

## Results and Discussion

Both accuracy and speed, i.e. finding the right structure consistently in biologically functional time, are important criteria in the protein folding problem. We measure accuracy of the local search in terms of *Success Rate* (*SR*), which is the proportion of total runs (20) per PRN where the hill climber found the unique global optimum within 1 million evaluations. Speed of the local search is accessed by *avg_evals*, which is the number of fitness function evaluations averaged over all runs with SR > 0.0 per PRN. Configuration *A* is considered more favorable to protein folding than configuration *B* if the local search algorithm performs better, i.e. achieves a significantly higher SR and a significantly lower avg_evals, on *A* than on *B*. A *configuration* refers to a combination of network topology and fitness function. Search performance is affected by network size, larger networks are in general expected to be either more difficult to optimize and/or require more function evaluations. To remove this size effect, search performances between two configurations are compared on the set of common networks with SR > 0.0. The largest p-value of the Shapiro-Wilk test for SR and avg_evals data is 0.03380715 and 1.072698e-07 respectively. This allows us to conclude, with at least 95% confidence, that both SR and avg_evals data are not normally distributed, and to use the Wilcoxon method (paired) to test SR and avg_evals data for significance. Following this procedure, the hypotheses in the following discussion are confirmed with at least 95% confidence.

## Experiment 1

The objective is to compare PRNs with lattices ("regular" graphs), i.e. how differences in their network topology affect search performance given the fitness function stated earlier. Lattice$V$ is a linear lattice with $V/2$ nearest neighbours to the left and to the right where possible ($V/2$ nodes at each of the two ends of the lattice chain will have fewer links than the rest of the nodes in the middle which will have $V$ links each). The GH64 PRNs share several network characteristics with Lattice4 and Lattice8. For networks having the same number of nodes, PRNs have the same link density as Lattice8, and similar clustering and positive degree assortativity levels as Lattice4 (Figures 9, 4 and 7 respectively).

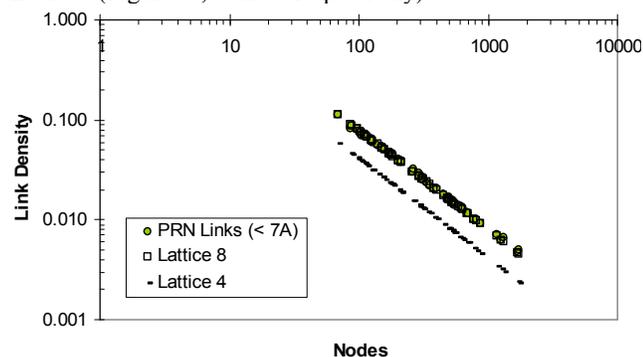

**Figure 9** Link density by size for GH64 proteins on a log-log plot, compared with the same for Lattice4 and Lattice8. Link density is the fraction of actual links out of all possible links, i.e. 2M / [N (N-1)].

Using the fitness function defined earlier, the hill climber performed better when the network topology is PRN than when it is Lattice8 (Table 1). However, both are outperformed by Lattice4 (Table 1), which has a significantly longer APL (Figure 6). Lattice4 also produced twice as many networks with SR > 0.5, and seems more conducive to larger networks than either Lattice8 or PRN (Table 2). Hence, shorter inter-nodal distances do not guarantee better search performances. The shorter APLs of Lattice8 and of PRN are the result of more links (Figure 9), and the fitness function is such that additional links can increase frustration (more on this point in Experiment 4). Furthermore, the regular connectivity of a lattice network probably does not produce suitably functional surface structures like those of proteins (but see Table 7).

**Table 1** Result summary for Experiment 1

| A | B | # | SR | avg_evals |
|---|---|---|---|---|
| Lattice8 | **PRN** | 49 | A = B | A > B |
| **Lattice4** | Lattice8 | 53 | A > B | A < B |
| **Lattice4** | PRN | 56 | A > B | A < B |

# is the number of networks with SR > 0.0, common to both configurations *A* and *B*. Configurations with better overall search performances are **bolded**. PRNs are from the GH64 dataset. Optimal strings for PRN and both Lattice4 and Lattice8 come from the DSSP output for proteins in the GH64 dataset (as explained earlier in the Method section).

## Experiment 2

The objective is to observe the effect of link randomization on search performance. By randomizing different sets of links in a PRN, it is possible to rearrange the links of a PRN such that the APL is significantly reduced (Figure 8). Further, by randomizing only the long-range links (randLE), it also possible to significantly reduce APLs of PRNs and maintain clustering coefficients and positive degree-degree assortativity at levels higher than would be in random graphs (Figure 8).

**Table 3** Result Summary for Experiment 2

| A | B | # | SR | avg_evals |
|---|---|---|---|---|
| randSE | randAll | 64 | A = B | A = B |
| PRN | **randLE** | 55 | A < B | A > B |
| PRN* | **randLE*** | 115 | A < B | A > B |

PRNs are from the GH64 and where indicated by *, the EVA132 dataset. randLE, randSE and randAll are as explained earlier, PRN networks produced by respectively randomizing only the Long-range links, only the Short-range links and all links. Optimal strings for all networks come from the DSSP output for GH64 proteins and where indicated by *, the DSSP output for EVA132 proteins.

Compared to PRN, randomizing all links (randAll) and randomizing short-range links (randSE) increased SR to almost 100%, with a considerable decrease in avg_evals (Table 2). There is no significant difference in terms of search performances between randSE and randAll (Table 3). But randAll and randSE networks lose much of the local organizational structure of PRN networks (Figure 8), and so probably do not produce suitably functional surface structures like those of proteins (see Table 7). What is more interesting is that randomizing long-range links (randLE) significantly improved search performance over PRN (Table 3) for both GH64 and EVA132 datasets. We revisit this point in Experiment 4, where an adjustment to the fitness function restores PRN to a favourable spot in the space of possible network configurations considered in this paper.

## Experiment 3

The objective is to compare the relative importance of short-range and long-range links to protein folding. There has been quite an evolution of thought in this area (Dill et al, 1995; Gō, 1983) and is by no means a settled issue. All three possible perspectives have been contemplated: (i) primacy of short-range interactions, (ii) primacy of long-range interactions, and (iii) non-dominance of either set of interactions.

Table 2 Key summary statistics for results obtained in Experiments 1, 2 and 3

| PRNs | Configuration | Median, Mean, Sd Success Rate (SR) | Proportion of networks with > 0.0 SR | Median avg_evals of networks with > 0.0 SR | Proportion of networks with > 0.5 SR | Median size of networks with > 0.5 SR |
|---|---|---|---|---|---|---|
| GH64 | Lattice4 | 0.6000, 0.5727, 0.3142 | 62/64 = 0.9688 | 6158 | 35/64 = 0.5469 | 211 |
|  | Lattice8 | 0.2500, 0.3289, 0.3190 | 53/64 = 0.8281 | 7245 | 15/64 = 0.2344 | 185 |
|  | PRN | 0.3250, 0.3383, 0.2569 | 56/64 = 0.8750 | 6264 | 14/64 = 0.2188 | 129 |
|  | randLE | 0.3750, 0.4000, 0.2772 | 62/64 = 0.9688 | 5126 | 24/64 = 0.3750 | 134 |
|  | randSE | 1.0000, 0.9977, 0.0139 | 64/64 = 1.0000 | 3609 | 64/64 = 1.0000 | 290 |
|  | randAll | 1.0000, 0.9945, 0.0220 | 64/64 = 1.0000 | 3541 | 64/64 = 1.0000 | 290 |
|  | onlySE | 0.0500, 0.1313, 0.2124 | 33/64 = 0.5156 | 4115 | 6/64 = 0.0938 | 134 |
|  | onlyLE | 1.0000, 0.9070, 0.1466 | 64/64 = 1.0000 | 4262 | 62/64 = 0.9688 | 276 |
|  | delay07 | 0.3500, 0.3734, 0.2619 | 58/64 = 0.9063 | 5703 | 20/64 = 0.3125 | 146 |
|  | delay08 | 0.3500, 0.3703, 0.2735 | 57/64 = 0.8906 | 5926 | 18/64 = 0.2813 | 146 |
|  | delay09 | 0.3500, 0.3789, 0.2823 | 56/64 = 0.8750 | 5733 | 21/64 = 0.3281 | 153 |
|  | delay10 | 0.0500, 0.1273, 0.2074 | 33/64 = 0.5156 | 4115 | 6/64 = 0.0938 | 134 |
| EVA132 | PRN | 0.2500, 0.3008, 0.2460 | 119/132 = 0.9015 | 9976 | 27/132 = 0.2045 | 145 |
|  | randLE | 0.3500, 0.3553, 0.2508 | 120/132 = 0.9091 | 10580 | 33/132 = 0.2500 | 226 |

The first column gives the source of the PRN and optimal string. **s**. Size of networks in the GH64 dataset has a median of 290 and is not normally distributed. The one-sample Kolmogorov-Smirnov two-sided test p-value = 2.220e-16. Size of networks in the EVA132 dataset has a median of 442 and is not normally distributed. The one-sample Kolmogorov-Smirnov two-sided test p-value is < 2.2e-16.

Compared with PRN, the use of only short-range links to guide the search (onlySE) reduced SR by 41% while using only long-range links (onlyLE) increased SR by 14% to almost 100% (Table 2). By examining both GH64 and EVA132 PRNs, we found that on average, only about 30% of all links in a PRN are long-range. The satisfaction of all links (constraints) in a PRN is necessary to achieve the globally maximal fitness value, and perfectly relaxed protein molecules as described by Gō (1983). Thus, the SR for onlyLE is quite remarkable and lends credence to the "primacy of long-range interactions" view (Dill et al, 1995). In both GH64 and EVA132 PRNs, long-range interactions implicate on average about 60% of all nodes in a PRN. However, it has been proposed that only 30% of residues are crucial for folding (Dill 1999, p.1169).

Proteins exist in 3D physical space. The possibility of a long-range link may depend on some prior sequence of events to bring distant nodes on the polypeptide chain into close spatial proximity with each other. Hence, there is some dependency between links. But long-range links are not mere corollaries to short-range links. Gō (1983) argues that "…folding cannot be a simple unidirectional sequence of events going from smaller to larger structures; long-range interactions also play a determining role in secondary structures and there should be feedback of logic between the levels of organization".

In our experiments, we observed that a slight delay in the use of long-range links to guide the hill climber significantly improved search performance (Table 4). In delayZ, the use of long-range links is delayed until the fraction of satisfied short-range constraints reaches Z/10. However, if long-range links are included only after all short-range links are satisfied, as in delay10, SR drops to levels similar to SR for onlySE (Table 2), i.e. it is as though long-range links are ignored in the search process completely. These results show that long-range links do help the satisfaction of short-range links in PRN (illustrating Go's point), but the involvement of long-range links in the search is more productive once some level of satisfaction (> 50%) in short-range links have occurred.

Table 4 Result Summary for Experiment 3

| A | B | # | SR | avg_evals |
|---|---|---|---|---|
| PRN | **delay07** | 54 | A < B | A > B |
| PRN | **delay08** | 54 | A < B | A > B |
| PRN | **delay09** | 53 | A < B | A = B |

# is the number of networks with SR > 0.0, common to both configurations *A* and *B*. Configurations with better overall search performances are **bolded**. PRNs are from the GH64 dataset. delay07, delay08 and delay09 are, as explained later in the text, PRN networks produced by delaying the consideration of long-range links when computing the fitness function until a proportion of short-range links are satisfied. Optimal strings for all networks come from the DSSP output for proteins in the GH64 dataset.

We collected the search points (strings) at the end of failed runs for one PRN (1wba) which failed fairly evenly under the different scenarios tested, and summarized their fitness values and Hamming distances from the global optimum. There is a negative correlation between fitness and distance. We find that runs which did not have enough opportunity to use long-range links to guide the search (onlySE, delay10) produced strings which are most distant and also least fit at the end (Figure 10). Even though the PRN strings are from failed runs, they are still more fit and closer to the global optimum than strings from delay10 or from onlySE.

## Experiment 4

The objective is to observe the effect of varying fitness contribution of links by link length on search performance, given the previous findings. To incorporate the outcome of experiment 3 into the optimization function, we added a weight factor into the *f* term for the experiments in this section as follows: $\sum_{i=0}^{N} g(s_i, \mathbf{s_1}) + \sum_{i=0}^{M} f(e_i, s, \mathbf{s}, \omega)$. There are three options to ω: (i) *eq* which assigns equal weight or fitness contribution to all links (this is the option used so far); (ii) *bh* which assigns more weight to links with shorter range; and (iii) *th* which assigns more weight to links with longer range. Let $e_i$ link nodes *j* and *k*, $d = |j - k|$ and $|s_j - s_k| = |\mathbf{s}_j - \mathbf{s}_k|$. If ω = *eq*, $f(e_i, s, \mathbf{s}, \omega) = 1$. If ω = *bh*, $f(e_i, s, \mathbf{s}, \omega) = 1/d$. If ω = *th*, $f(e_i, s, \mathbf{s}, \omega) = d$. However, ω = *th* produced 0.0 SR for GH64 PRNs, and therefore is clearly not a viable fitness function (This outcome is not contradictory to the onlyLE result in Experiment 3 because there, the fitness function assigns equal weights to all links). As such we restrict the following discussion to *bh* and *eq* options.

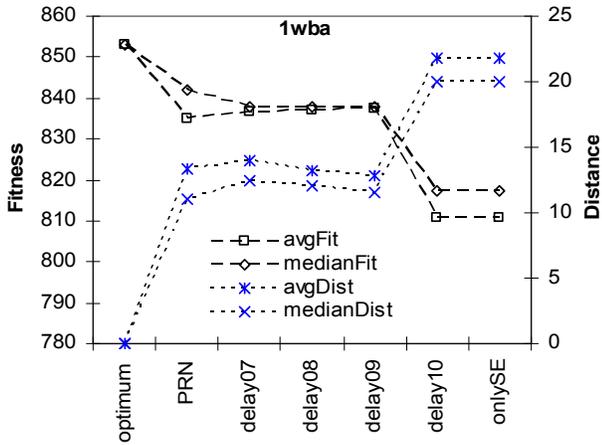

**Figure 10** Summarized fitness and hamming distance from global optimum (optimum) of strings produced by failed runs for 1wba PRN. Number of strings in PRN, delay07, delay08, delay09, delay10 and onlySE are 18, 16, 17, 18, 20 and 20 respectively.

Compared with the *eq* option (Table 1), the *bh* option significantly improved search performance when PRN is used as the underlying network (Table 5). For both GH64 and EVA132, the number of networks with SR > 0.5 increased by at least 2.5 times (36/14 and 77/26), and there is also an increase of at least 55% in the median size of networks with SR > 0.5 (Table 6). Putting more weight on short-range than long-range links introduces a bias towards the satisfaction of short-range links and is akin to putting a delay on long-range links as we did in the delayZ runs where Z < 1.0 (Experiment 3). However, with the *bh* option, both sets of interactions are present right from the start, so they have more interplay opportunities. And from the results just discussed, there appears to be a payoff to this. Also, PRN_bh produced significantly better search performance than delay07_eq (Table 5). Hence, the fitness function with the *bh* option appears to be more compatible to the suitability of PRNs to protein folding.

When ω = *bh*, as in Experiment 1, search performance is still better when the underlying network topology is PRN than when it is Lattice8 (Table 5). However, unlike Experiment 2, search performance is no longer significantly better with randLE than with PRN. For both GH64 and EVA132, when ω = *bh*, there is no significant difference between PRN and randLE in terms of search performance (Table 5). Hence, the shorter APLs that randLE can produce (Figure 8) do not confer a search advantage. With this finding, we observe as in Experiment 1, that shorter APLs do not necessarily guarantee better search performance. However, in this case, no additional links are involved.

**Table 5** Result Summary for Experiment 4

| A | B | # | SR | avg_evals |
|---|---|---|---|---|
| **PRN_bh** | PRN | 56 | A > B | A < B |
| **PRN*_bh** | PRN* | 116 | A > B | A < B |
| **PRN_bh** | delay07 | 58 | A > B | A < B |
| **PRN_bh** | Lattice8_bh | 56 | A > B | A < B |
| **randLE_bh** | Lattice8_bh | 56 | A > B | A < B |
| PRN_bh | randLE_bh | 61 | A = B | A = B |
| PRN*_bh | randLE*_bh | 122 | A = B | A = B |

The '_bh' suffix is used to mark configurations which use the *bh* option; otherwise the default *eq* option is used. PRNs are from the GH64 and where indicated by *, the EVA132 dataset. Optimal strings for all networks come from the DSSP output for GH64 proteins and where indicated by *, the DSSP output for EVA132 proteins.

## Summary and Conclusion

In experiment 1, we observed that PRN is a more favourable network topology than Lattice8 for protein folding, but that shorter average path lengths (APLs) need not imply better search performance. In experiment 2, we reported that randomizing long-range links of protein residue networks (randLE) significantly improved search performance over (non-randomized) PRNs. In experiment 3, we found that long-range links play an important role to global optimization and that adding a delay to the involvement of long-range links in the search improved search performance. In experiment 4, we use a modified fitness function which assigns more fitness contribution to shorter links and found that indeed PRN is a more favourable network topology than randLE for protein folding even though PRN has a significantly longer APL than randLE.

Shorter APLs in PRNs imply more compactness in native state proteins. That PRNs do not have minimal or at least shorter APLs than they do agrees with the notion that native state proteins are not in the most compact conformation possible (Dill et al, 1995 p. 568).

With ω = *bh* in Experiment 4, PRNs appear to occupy a sweet spot between complete order and total randomness PRNs outperformed Lattice8 in terms of search performance, and random graphs (e.g. randSE and randAll networks) are unlikely to produce viable protein structures (Vendruscolo *et al* 1999). What about randLE which produced a comparable search performance to PRN? randLE represent less plausible 3D structures than PRNs, but more plausible than randSE or randAll (Table 7).

Randomization of long-range links in PRNs (randLE) was performed while keeping node degree of PRNs invariant. Hence, our experiments also suggest that there can be explanations, other than the popular excluded volume argument, beneath the topological limits of PRNs. For

Table 6 Key summary statistics for results obtained in Experiment 4

| PRNs | Configuration | Median, Mean, Sd Success Rate (SR) | Proportion of networks with > 0.0 SR | Median avg_evals of networks with > 0.0 SR | Proportion of networks with > 0.5 SR | Median size of networks with > 0.5 SR |
|---|---|---|---|---|---|---|
| GH64 | Lattice8_bh | 0.5000, 0.5023, 0.3281 | 56/64 = 0.8750 | 4624 | 31/64 = 0.4844 | 203 |
|  | PRN_bh | 0.5750, 0.5750, 0.3176 | 62/64 = 0.9688 | 4947 | 36/64 = 0.5625 | 208 |
|  | randLE_bh | 0.5750, 0.5648, 0.3172 | 62/64 = 0.9688 | 5336 | 34/64 = 0.5313 | 208 |
|  | randSE_bh | 1.0000, 0.8773, 0.2047 | 64/64 = 1.0000 | 3283 | 57/64 = 0.8906 | 213 |
|  | randAll_bh | 1.0000, 0.8352, 0.2426 | 64/64 = 1.0000 | 3344 | 54/64 = 0.8438 | 212 |
| EVA132 | PRN*_bh | 0.6000, 0.5614, 0.3176 | 124/132 = 0.9394 | 9219 | 77/132 = 0.5833 | 226 |
|  | randLE*_bh | 0.6000, 0.5492, 0.3197 | 122/132 = 0.9242 | 9373 | 76/132 = 0.5758 | 274 |

The first column gives the source of the PRN and optimal string **s**. The '*' indicates the use of the EVA132 dataset. The '_bh' suffix indicates the use of the *bh* option in the fitness function.

Table 7 FT-COMAR results for five random PRNs in GH64

| PID | N | Lat8 | PRN | randLE | randSE | randAll |
|---|---|---|---|---|---|---|
| 153L | 185 | 0 | 0 | 196 | 1501 | 1543 |
| 1arb | 263 | 0 | 0 | 1319 | 2278 | 2221 |
| 1cjl | 307 | 0 | 144 | 1132 | 2802 | 2870 |
| 1rpx | 690 | 0 | 355 | 3208 | 6496 | 6589 |
| 1psd | 808 | 0 | 651 | 4975 | 7559 | 7772 |

FT-COMAR predicts a plausible 3D construction of a given contact map and threshold, and reports the Hamming distance between the given contact map and the contact map of the predicted structure. Hence, a larger value in this table implies that the given contact map is less plausible as a 3D structure. FT-COMAR works better for thresholds larger than the one we use, i.e. 7 Angstrom, and this affects the results for larger PRNs. Nonetheless, the results still favor PRN over randLE.

instance, both local (high clustering and positive assortativity) and global (short average path length) characteristics of PRN seem necessary to create favorable conditions for protein folding.

Finally, it could be worthwhile, to both protein folding studies and systems organization in general, to understand how the lower and higher levels (i.e. short-range and long-range links respectively) in proteins cooperate to create mutual satisfaction without either level necessarily dominating the process.